\documentclass[12pt]{article} 
\usepackage[sectionbib]{natbib}
\usepackage{array,epsfig,fancyheadings,rotating}
\usepackage[]{hyperref}  
\usepackage{sectsty, secdot}
\sectionfont{\fontsize{12}{14pt plus.8pt minus .6pt}\selectfont}
\renewcommand{\theequation}{\thesection\arabic{equation}}
\subsectionfont{\fontsize{12}{14pt plus.8pt minus .6pt}\selectfont}

\usepackage{xr}
\usepackage{amsmath}
\usepackage{amssymb}
\usepackage{amsfonts}
\usepackage{multirow}
\usepackage{amsthm}
\usepackage{xcolor}

\allowdisplaybreaks

\usepackage{geometry}
\usepackage{booktabs}
\usepackage{float}

\bibpunct{(}{)}{,}{autheryear}{,}{;}

\newcommand*{\cov}{\mathrm{cov}}

\newcommand{\T}{\mathrm{\scriptscriptstyle T}}
\newcommand*{\mi}{\mathrm{MI}}

\newcommand*{\obs}{\mathrm{obs}}

\newcommand*{\N}{\mathcal{N}}
\newcommand*{\F}{\mathcal{F}}
\newcommand*{\J}{\mathcal{J}}
\newcommand*{\de}{\mathrm{d}}
\newcommand*{\reg}{\mathrm{reg}}

\newcommand{\ipw}{\mathrm{IPW}}
\newcommand{\aipw}{\mathrm{AIPW}}

\newcommand{\mat}{\mathrm{mat}}

\newcommand{\beq}{ \begin{equation*}}
\newcommand{\eeq}{ \end{equation*}}
\newcommand{\beqn}{ \begin{eqnarray}}
\newcommand{\eeqn}{ \end{eqnarray}}
\newtheorem{corollary}{Corollary}\newtheorem{theorem}{Theorem}\newtheorem{remark}{Remark}

\newtheorem{example}{Example}\newtheorem{assumption}{Assumption}

\newcommand*{\pr}{\mathbb{P}}
\newcommand*{\E}{\mathbb{E}}
\newcommand*{\var}{\mathbb{V}}

\begin{document}

\renewcommand{\baselinestretch}{2}

\markright{ \hbox{\footnotesize\rm Statistica Sinica
}\hfill\\[-13pt]
\hbox{\footnotesize\rm
}\hfill }

\markboth{\hfill{\footnotesize\rm QIAN GUAN AND SHU YANG} \hfill}
{\hfill {\footnotesize\rm A unified inference framework for multiple imputation using
		martingales} \hfill}

\renewcommand{\thefootnote}{}
$\ $\par

	
\fontsize{12}{14pt plus.8pt minus .6pt}\selectfont \vspace{0.8pc}
\centerline{\large\bf A unified inference framework for multiple imputation using
	martingales}
\vspace{.4cm} 
\centerline{Qian Guan and Shu Yang}
\centerline{\it Department of Statistics, North Carolina State University}
\vspace{.55cm} \fontsize{9}{11.5pt plus.8pt minus.6pt}\selectfont
	

\begin{quotation}
\noindent {\it Abstract:} Multiple imputation is widely used to handle missing data. Although Rubin's combining rule is simple, it is not clear whether or not the standard multiple imputation inference is consistent when coupled with the commonly-used full sample estimators. This article establishes a unified martingale representation of multiple imputation for a wide class of asymptotically linear full sample estimators. This representation invokes the wild bootstrap inference to provide consistent variance estimation \textcolor{black}{under the correct specification of the imputation models}. As a motivating application, we illustrate the proposed method to estimate the average causal effect (ACE) with partially observed confounders in causal inference. Our framework applies to asymptotically linear ACE estimators, including the regression imputation, weighting, and matching estimators. We extend to the scenarios when both outcome and confounders are subject to missingness and when the data are missing not at random.

\vspace{9pt}
\noindent {\it Key words and phrases:}
Causality; Congeniality; Martingale representation;
Influence function; Weighted bootstrap. 
\par
\end{quotation}\par

\def\thefigure{\arabic{figure}}
\def\thetable{\arabic{table}}

\renewcommand{\theequation}{\thesection.\arabic{equation}}

\fontsize{12}{14pt plus.8pt minus .6pt}\selectfont

	\section{Introduction\label{sec:intro}}
	
	Missing data are ubiquitous in practice. A widely-used approach to
	handle incomplete/missing data is multiple imputation (MI). The National
	Research Council has recommended MI as one of its preferred approaches
	to addressing missing data \citep{national2010prevention}. The idea
	of MI is to fill the missing values multiple times by sampling from
	the posterior predictive distribution of the missing values given
	the observed values. Then, full sample analyses can be applied straightforwardly
	to the imputed data sets, and these multiple results are summarized
	by an easy-to-implement combining rule for inference \citep{rubin1987multiple}.
	MI can provide valid frequentist inferences in various applications
	(e.g., \citealp{clogg1991multiple}). On the other hand, many authors
	have found that Rubin's variance estimator is not always consistent
	(e.g., \citealp{fay1992inferences}, \citealp{kott1995paradox}, \citealp{fay1996alternative},
	\citealp{binder1996frequency}, \citealp{wang1998large}, \citealp{robins2000inference},
	\citealp{nielsen2003proper} and \citealp{kim2006bias}). To ensure
	the validity of Rubin's variance estimation, imputations must be proper
	\citep{rubin1987multiple}. A sufficient condition for proper imputation
	is the congeniality condition of \citet{meng1994multiple}, imposed
	on both the imputation model and the subsequent full sample analysis.
	Even with a correctly specified imputation model, \citet{yang2016mi}
	showed that MI is not necessarily congenial for the method of moments
	estimation, so common statistical procedures can be incompatible with
	MI. Given the popularity of MI in practice, it is important to develop
	a valid inference procedure for utilizing MI in statistical inference.
	
	As a motivating application, we focus on causal inference with partially
	observed confounders. Causal inference is a central goal in many disciplines,
	such as medicine, econometrics, political and social sciences. When
	all confounders that influence both treatment and outcome are observed,
	the average causal effect (ACE) of the treatment is identifiable \citep{imbens2015causal}.
	The literature has proposed many ACE estimators, such as regression
	imputation (\citealp{hahn1998role,heckman1997matching}), (augmented)
	propensity score weighting (\citealp{horvitz1952generalization,rosenbaum1983central,robins1994estimation,bang2005doubly,cao2009improving})
	and matching (\citealp{rosenbaum1989optimal,stuart2010matching,abadie2016matching})
	to adjust for confounders. Previous works have used MI for causal
	inference with partially observed confounders, e.g., \citet{qu2009propensity},
	\citet{crowe2010comparison}, \citet{mitra2011estimating}, and \citet{seaman2014inverse}.
	Given that many full sample estimators are available for estimating
	the ACE, the validity of Rubin's variance estimator using these full
	sample estimators for causal inference is  largely
	unexplored.
	
	In this article, we establish a novel martingale representation of
	MI for a general class of asymptotically linear full sample estimators
	\textcolor{black}{under the correct specification of the imputation models}.
	Our key insight is that the MI estimator is intrinsically created
	in a sequential manner: first, the posterior samples of parameters
	are drawn from the posterior distribution, which is asymptotically
	equivalent to the sampling distribution of the maximum likelihood
	estimator based on the Bernstein-von Mises theorem (\citealp{van1998asymptotic};
	Chapter 10); second, the posterior predictive samples of the missing
	data are drawn conditioned on the observed data. This conceptualization
	leads to an asymptotically linear expression of the MI estimator in
	terms of a sequence of random variables that have conditional mean
	zero given the sigma algebra generated from the preceding variables
	(i.e., a martingale representation). The martingale representation
	invokes the wild/weighted bootstrap procedure \citep{wu1986jackknife,liu1988bootstrap}
	that provides valid variance estimation and inference regardless of
	which full sample estimator is adopted in MI. 
	
	We show the asymptotic validity of our proposed bootstrap inference
	method for the MI estimator using the martingale central limit theory
	\citep{hall1980martingale} and the asymptotic property of weighted
	sampling of martingale difference arrays \citep{pauly2011weighted}.
	Although the validity of the proposed method is based on the asymptotic
	results as the sample size goes to infinity, the simulation results
	demonstrate that it performs well for finite samples. It is worthwhile
	to compare the proposed method with the improper MI approach proposed
	by \citet{wang1998large} and \citet{robins2000inference}. The idea
	of improper MI is to use Monte Carlo imputation as a tool to compute
	the maximum likelihood estimator and therefore, it requires the imputation
	size $m$ to be large in order to reduce the Monte Carlo error. In
	contrast, our proposed method allows the imputation size $m$ to be
	fixed at a small value. This property is appealing for releasing multiply
	imputed datasets for public usage. Moreover, improper MI can only
	deal with regular estimators but not non-regular estimators such as
	the matching estimators. The proposed method can be applied to a wide
	range of the ACE estimators adopted in MI, including the outcome regression,
	weighting, and matching estimators. Indeed, the simulation studies
	indicate that Rubin's variance estimator overestimates the variance
	for the IPW and matching estimators because these two estimators are
	not self-efficient (\citealp{meng1994multiple,xie2017dissecting}),
	while the proposed variance estimation procedure is consistent for
	all types of estimators.
	
	Importantly, our framework can easily accommodate the scenarios when
	both outcome and confounders have missing values and when the missing
	data are missing not at random. In the former case, we only need to
	add the imputation step for the missing outcomes. In the latter case,
	we only need to modify the imputation model by further considering
	the missing data probability model in the data likelihood function.
	Our research is likely to bridge the advantages of MI and its wide
	applications in causal inference and missing data analyses.
	
	The rest of the paper is organized as follows. Section \ref{sec:background}
	introduces general asymptotically linear estimators and illustrates
	with common estimators in causal inference. Section \ref{sec:Multiple-imputation}
	describes the general MI to fill in missing values that facilitate
	full sample estimators. Section \ref{sec:Martingale} presents the
	martingale representation for the MI estimators and the wild bootstrap
	inference procedure and establishes its validity. Section \ref{sec:Extension}
	extends the proposed method to the scenario with other causal estimands,
	the scenario where both outcome and the confounders have missing values
	and the scenario where the confounders are missing not at random.
	In Section \ref{sec:sim}, we evaluate the finite sample performance
	of the proposed method using simulation studies. In Section \ref{sec:realdata},
	we apply the proposed wild bootstrap inference method to a U.S. National
	Health and Nutrition Examination Survey data. Section \ref{sec:conclude}
	concludes.
	
	\section{Background\label{sec:background}}
	
	\subsection{General setup \label{subsec:General-setup}}
	
	We introduce a general setup and illustrate it with common estimators
	of the ACE in causal inference. Suppose we observe $n$ independent
	and identically distributed (i.i.d.) samples $\mathbf{L}=\{L_{i}:i=1,\ldots,n\}$
	governed by the distribution $\pr(L)$. We are interested in inference
	about the target parameter, a functional of the observed data distribution,
	$\tau=\tau(\pr)$, e.g., the mean of the distribution $\pr$. For simplicity of presentation, we assume $\tau$ to be a one-dimensional parameter. An extension to a multi-dimensional parameter is feasible at the cost of heavier notation. Let
	$\hat{\tau}_{n}$ denote a generic estimator of $\tau$. We focus
	on the class of asymptotically linear estimators. This class of estimators
	includes the common regular and asymptotically linear{{}
	}(RAL) estimators, which can be expressed by 
	\begin{equation}
	\hat{\tau}_{n}-\tau=\frac{1}{n}\sum_{i=1}^{n}\psi(L_{i})+o_{\pr}(n^{-1/2}),\label{eq:linear form}
	\end{equation}
	where $\{\psi(L_{i}):i=1,\ldots,n\}$ are i.i.d with $\E\{\psi(L_{i})\}=0$
	and $\E\{\psi(L_{i})^{2}\}<\infty$. The random variable $\psi(L_{i})$
	is called the influence function of $\hat{\tau}_{n}$ and captures
	the first-order asymptotic behavior of $\hat{\tau}_{n}$ \citep{bickel1993efficient}.
	Regarding regularity conditions, see, e.g., \citet{newey1990semiparametric}.
	For a given estimator, upon identifying its influence function, we
	can characterize the asymptotic distribution and construct corresponding
	confidence intervals for the target parameter. The class of estimators
	also includes possibly non-regular asymptotically linear estimators,
	which can be expressed by 
	\begin{equation}
	\hat{\tau}_{n}-\tau=\frac{1}{n}\sum_{i=1}^{n}\psi_{i}(\mathbf{L})+o_{\pr}(n^{-1/2}),\label{eq:non-ral}
	\end{equation}
	where the individual component $\psi_{i}(\mathbf{L})$ may depend
	on the full sample and therefore is not i.i.d, but satisfies $\E\{\psi_{i}(\mathbf{L})\}=0$
	and $\E\{\psi_{i}(\mathbf{L})^{2}\}<\infty$. 
	 \textcolor{black}{The matching estimator
	is an example as we illustrate later. For simplicity,
	we also call $\psi_{i}(\mathbf{L})$ the influence function of $\hat{\tau}_{n}$.}
	
	\subsection{Motivating application: estimating average causal effects}
	
	We elucidate the general framework with an application of estimating
	the ACE. Let $X$ be a vector of $p$-dimensional
	covariates, $A\in\{$$0,1\}$ be a binary treatment, with $0$ and
	$1$ being the labels for control and active treatments, respectively,
	and $Y$ be the outcome of interest. Suppose we observe $n$ i.i.d.
	samples $\mathbf{L}=\{L_{i}=(A_{i},X_{i},Y_{i}):i=1,\ldots,n\}$.
	
	Following \citet{splawa1990application} and \citet{rubin1974estimating},
	we use the potential outcomes framework to formulate the causal parameter
	of interest. Under the Stable Unit Treatment Value assumption \citep{rubin1980comment},
	for each level of treatment $a$, there exists a potential outcome
	$Y(a)$, representing the outcome had the unit, possibly contrary
	to the fact, been given treatment $a$. We make the causal consistency
	assumption that links the observed outcome with the potential outcomes;
	i.e., the observed outcome $Y$ is the potential outcome $Y(A)$ under
	the actual treatment. We focus on estimating the ACE $\tau=\E\{Y(1)-Y(0)\}$.
	Our methodology applies to a broader class of causal estimands in
	\citet{li2018balancing}; we discuss the extension to other causal
	estimands in Section \ref{subsec:Other-esimands}. For simplicity
	of exposition, denote 
	\[
	\mu_{a}(X)=\E\{Y(a)\mid X\}\ \text{ and }\ e(X)=\pr(A=1\mid X),
	\]
	where $\mu_{a}(X)$ is an outcome mean function for $a=0,1$ and $e(X)$
	is the propensity score.
	

It is well known that under the common assumptions in the causal inference literature, \textcolor{black}{including the treatment ignorability and overlap assumptions (Assumptions \ref{asump-ignorable} and \ref{asump-overlap} in the supplementary material),} the ACE can be identified by various important estimators
that are widely used in practice, including outcome regression, augmented/inverse
probability weighting (AIPW/IPW), or matching. See \citet{imbens2004nonparametric}
and \citet{rosenbaum2002observational} for surveys of these estimators.
These common estimators are asymptotically linear and belong to the
class of estimators in our general setup. We review these estimators below and identify their
influence functions in the supplementary material.

\textcolor{black}{
The common estimators require correct specifications of different
parts of the observed data distribution, including the outcome model
and propensity score.}
\begin{assumption}[Outcome model] \label{asump outcome}
	The parametric model $\mu_{a}(X;\beta_{a})$ is a correct specification
	for $\mu_{a}(X)$, for $a=0,1$; i.e., $\mu_{a}(X)=\mu_{a}(X;\beta_{a}^{*})$,
	where $\beta_{a}^{*}$ is the true model parameter.
\end{assumption}

\begin{assumption}[Propensity score model] \label{asump ps}
	
	The parametric model $e(X;\alpha)$ is a correct specification for
	$e(X)$; i.e., $e(X)=e(X;\alpha^{*})$, where $\alpha^{*}$ is the
	true model parameter.
	
\end{assumption}

	\begin{example}\label{example:reg}
		\textcolor{black}{
		The outcome regression estimator is $\hat{\tau}_{n,\reg}=n^{-1}\sum_{i=1}^{n}\tau_{\reg,i}$,
		where 
		\begin{equation}
		\tau_{\reg,i}=\mu_{1}(X_{i};\hat{\beta}_{1})-\mu_{0}(X_{i};\hat{\beta}_{0}).\label{eq:regression}
		\end{equation}
	}
	\end{example}
	\begin{example}\label{example:ipw}
		\textcolor{black}{
		The IPW estimator is $\hat{\tau}_{n,\ipw}=n^{-1}\sum_{i=1}^{n}\tau_{\ipw,i},$
		where
		\begin{equation}
		\tau_{\ipw,i}=\frac{A_{i}Y_{i}}{e(X_{i};\hat{\alpha})}-\frac{(1-A_{i})Y_{i}}{1-e(X_{i};\hat{\alpha})}.\label{eq:ipw}
		\end{equation}
	}
	\end{example}
	
	
	\begin{example}\label{example:aipw}
		\textcolor{black}{
		The AIPW estimator i{\small{}s $\hat{\tau}_{n,\aipw}=n^{-1}\sum_{i=1}^{n}\tau_{\aipw,i}$},
		where
		\begin{multline}
		\tau_{\aipw,i}=\frac{A_{i}Y_{i}}{e(X_{i};\hat{\alpha})}+\left\{ 1-\frac{A_{i}}{e(X_{i};\hat{\alpha})}\right\} \mu_{1}(X_{i};\hat{\beta}_{1})\\
		-\frac{(1-A_{i})Y_{i}}{1-e(X_{i};\hat{\alpha})}-\left\{ 1-\frac{1-A_{i}}{1-e(X_{i};\hat{\alpha})}\right\} \mu_{0}(X_{i};\hat{\beta}_{0}).\label{eq:aipw}
		\end{multline}
	}
	\end{example}
	
	
	\begin{example}[Matching]\label{example:mat}
		\textcolor{black}{
		For unit $i$, denote the imputed potential outcomes as 
		\[
		\hat{Y}_{i}(1)=\begin{cases}
		M^{-1}\sum_{j\in\J_{X}(i)}Y_{j} & \text{if }A_{i}=0,\\
		Y_{i} & \text{if }A_{i}=1,
		\end{cases}\quad\hat{Y}_{i}(0)=\begin{cases}
		Y_{i} & \text{if }A_{i}=0,\\
		M^{-1}\sum_{j\in\J_{X}(i)}Y_{j} & \text{if }A_{i}=1.
		\end{cases}
		\]
		The matching estimator of $\tau$ is 
		\begin{equation}
		\hat{\tau}_{n,\mat}^{(0)}=\frac{1}{n}\sum_{i=1}^{n}\{\hat{Y}_{i}(1)-\hat{Y}_{i}(0)\}=\frac{1}{n}\sum_{i=1}^{n}(2A_{i}-1)\left(Y_{i}-M^{-1}\sum_{l\in\J_{X}(i)}Y_{l}\right).\label{eq:matching est}
		\end{equation}
	where $M$ ($M\geq1$) is the number of matches and $\J_{X}(i)$ is the index
	set of the nearest $M$ neighbors for unit $i$ in its opposite treatment
	group based on the matching variable $X$.
	}
	\end{example}

The above estimators are asymptotically linear with the influence functions given in the supplementary material. 
	
	\section{Multiple Imputation to Deal with Missing Values \label{sec:Multiple-imputation}}
	
	\subsection{General multiple imputation \label{subsec:General-multiple-imputation}}
	
	Continuing with the general setup in Section \ref{subsec:General-setup},
	we now consider the case where $L$ is $q$-dimensional and $L=(L_{[1]},\ldots,L_{[q]})$
	contains missing values. Let $R=(R_{[1]},\ldots,R_{[q]})$ be the
	vector of missing indicators such that $R_{[j]}=1$ if the $j$th
	component $L_{[j]}$ is observed and $0$ if it is missing. Also,
	let $1_{q}$ denote the $q$-vector of $1$'s. We write $L=(L_{R},L_{\overline{R}})$,
	where $L_{R}$ and $L_{\overline{R}}$ represent the observed and
	missing parts of $L$, respectively. This notation depends on the
	missingness pattern; e.g., if $R_{[1]}=1$ and $R_{[j]}=0$ for $j=2,\ldots,q$,
	then $L_{R}=L_{[1]}$ and $L_{\overline{R}}=(L_{[2]},\ldots,L_{[q]})$.
	With missing values in $L$, the full sample estimator $\hat{\tau}_{n}$
	is not feasible to calculate.
	
	To facilitate applying a full sample estimator, MI creates multiple
	complete data sets by filling in missing values. Assume unit $i$
	has the complete data $Z_{i}=(L_{i},R_{i})$ and the observed data
	$Z_{\obs,i}=(L_{R_{i},i},R_{i})$. Denote $\mathbf{Z}=(Z_{1},\ldots,Z_{n})$
	and $\mathbf{Z}_{\obs}=(Z_{\obs,1},\ldots,Z_{\obs,n})$. Assume that
	the observed data likelihood is $f(\mathbf{Z}_{\obs};\theta)$ with
	the true parameter value $\theta_{0}$. The MI procedure proceeds
	as follows. 
	\begin{description}
		\item [{Step$\ $MI-1.}] Create $m$ complete data sets by filling in missing
		values with imputed values generated from the posterior predictive
		distribution. Specifically, to create the $j$th imputed data set,
		first generate $\theta^{*(j)}$ from the posterior distribution $p(\theta\mid\mathbf{Z}_{\obs})$,
		and then generate $L_{\overline{R}_{i},i}^{*(j)}$ from $f(L_{\overline{R}_{i},i}\mid Z_{\obs,i};\theta^{*(j)})$
		for each missing $L_{\overline{R}_{i},i}$. 
		\item [{Step$\ $MI-2.}] Apply a full sample estimator of $\tau$ to each
		imputed data set. Let $\hat{\tau}^{(j)}$ be the estimator applied
		to the $j$th imputed data set, and $\hat{V}^{(j)}$ be the full sample
		variance estimator for $\hat{\tau}^{(j)}$. 
		\item [{Step$\ $MI-3.}] Use Rubin's combining rule to summarize the results
		from the multiple imputed data sets. The MI estimator of $\tau$ is
		$\hat{\tau}_{\mi}=m^{-1}\sum_{j=1}^{m}\hat{\tau}^{(j)}$, and Rubin's
		variance estimator is 
		\begin{equation}
		\hat{V}_{\mi}(\hat{\tau}_{\mi})=W_{m}+(1+m^{-1})B_{m},\label{eq:rubin-1}
		\end{equation}
		where $W_{m}=m^{-1}\sum_{j=1}^{m}\hat{V}^{(j)}$ and $B_{m}=(m-1)^{-1}\sum_{j=1}^{m}(\hat{\tau}^{(j)}-\hat{\tau}_{\mi})^{2}$. 
	\end{description}

\begin{remark}\label{rem:step}
In Step MI-1, as an anonymous referee pointed out, the full/observed data likelihood has to be specified and fitted for multiple imputation, which can be challenging in the presence of several, if not many, variables. In practice, we suggest specifying the full data likelihood as a product of a sequence of conditional models of one variable given the proceeding variables, allowing model flexibility for each variable (e.g., the error distribution matches the variable type — logistic for a binary variable). Also, model diagnosis can be carried out after imputation to assess goodness-of-fit. See the real-data application in Section \ref{sec:realdata} for an example. 	
\end{remark}
	
	\subsection{CI in the presence of confounders missing at random}
	
	We elucidate our method in the motivating application of estimating
	the ACE by assuming the confounders are missing at random (MAR) in
	the sense of \citet{rubin1976inference}. Extensions to settings with
	missing outcomes and different missingness mechanisms are provided
	in Section \ref{sec:Extension}. We now consider the case where $X=(X_{[1]},\ldots,X_{[p]})$,
	a $p$-dimensional vector, contains missing values. Accordingly, let
	$R_{X}=(R_{[1]},\ldots,R_{[p]})$ be the vector of missing indicators
	such that $R_{[j]}=1$ if the $j$th component $X_{[j]}$ is observed
	and $0$ if it is missing. We write $X=(X_{R_{X}},X_{\overline{R}_{X}})$,
	where $X_{R_{X}}$ and $X_{\overline{R}_{X}}$ represent the observed
	and missing parts of $X$, respectively. With missing values in $X$,
	the aforementioned full sample estimators (\ref{eq:regression})--(\ref{eq:matching est})
	are not feasible to calculate. Estimation of the ACE requires further
	assumptions. Following most of the empirical literature, we impose
	the MAR assumption.
	
	\begin{assumption}[Missingness at random]\label{assump:MAR}We
		have $X_{\overline{R}_{X}}\perp\!\!\!\perp R_{X}\mid Z_{\obs}.$
		
	\end{assumption}
	
	Assumption \ref{assump:MAR} holds if the observed data capture all
	the information related to missingness. Under Assumption \ref{assump:MAR},
	$f(A_{i},X_{i},Y_{i},R_{Xi};\theta)=f(A_{i},X_{R_{Xi},i},Y_{i},R_{Xi};\theta)$
	$f(X_{\overline{R}_{Xi},i}|A_{i},X_{R_{Xi},i},$ $Y_{i},R_{Xi}=1_{p};\theta)$
	is identifiable, which justifies the likelihood-based or Bayesian
	inference. Moreover, {by Bayes rule, the posterior
		distribution of the missing data can be expressed as} 
	\begin{eqnarray}
	&  & f(X_{\overline{R}_{Xi},i}\mid A_{i},X_{R_{Xi},i},Y_{i},R_{Xi};\theta^{*(j)})\propto f(A_{i},X_{\overline{R}_{Xi},i},X_{R_{Xi},i},Y_{i},R_{Xi};\theta^{*(j)})\nonumber \\
	& = & f(R_{Xi}\mid Y_{i},X_{R_{Xi},i},X_{\overline{R}_{Xi},i},A_{i};\theta^{*(j)})f(Y_{i},X_{R_{Xi},i},X_{\overline{R}_{Xi},i},A_{i};\theta^{*(j)})\nonumber \\
	& \propto & f(Y_{i},X_{R_{Xi},i},X_{\overline{R}_{Xi},i},A_{i};\theta^{*(j)})\label{eq:mar}\\
	& \propto & f(Y_{i}\mid X_{R_{Xi},i},X_{\overline{R}_{Xi},i},A_{i};\theta^{*(j)})f(A_{i}\mid X_{R_{Xi},i},X_{\overline{R}_{Xi},i};\theta^{*(j)})f(X_{\overline{R}_{Xi},i}\mid X_{R_{Xi},i};\theta^{*(j)}),\nonumber 
	\end{eqnarray}
	where (\ref{eq:mar}) follows because $f(R_{Xi}\mid Y_{i},X_{R_{Xi},i},X_{\overline{R}_{Xi},i},A_{i};\theta^{*(j)})=f(R_{Xi}\mid Y_{i},X_{R_{Xi},i},A_{i};\theta^{*(j)})$
	by Assumption \ref{assump:MAR}. The MI procedure proceeds with the
	imputation model for $X_{\overline{R}_{Xi},i}$, which does not depend
	on the missingness pattern probability for $R_{Xi}$.
	
	\subsection{Issue of standard inference with MI}
	
	The variance of the MI estimator can be decomposed to 
	\[
	\var(\hat{\tau}_{\mi})=\var(\hat{\tau}_{n})+\var(\hat{\tau}_{\mi}-\hat{\tau}_{n})+2\cov(\hat{\tau}_{\mi}-\hat{\tau}_{n},\hat{\tau}_{n}),
	\]
	In Rubin's variance estimator (\ref{eq:rubin-1}), $W_{m}$ estimates
	the within-imputation variance $\var(\hat{\tau}_{n})$, and $(1+m^{-1})B_{m}$
	estimates the between-imputation variance $\var(\hat{\tau}_{\mi}-\hat{\tau}_{n})$.
	However, it ignores the covariance between $\hat{\tau}_{\mi}-\hat{\tau}_{n}$
	and $\hat{\tau}_{n}$. Rubin's variance estimator is asymptotically
	unbiased only under the congeniality condition \citep{meng1994multiple},
	i.e., $\cov(\hat{\tau}_{\mi}-\hat{\tau}_{n},\hat{\tau}_{n})=o(1)$.
	Therefore, Rubin's variance estimator using the different full sample
	estimator $\hat{\tau}_{n}$ may be inconsistent.
	
	For illustration, we conduct a numerical experiment to assess the
	congeniality condition for the outcome regression, IPW, AIPW and matching
	estimators of the ACE. The data generating mechanism is described
	in scenario (a) in Section \ref{sec:sim}. For each simulated data
	set, we compute the full sample point estimators $\hat{\tau}_{n}$
	assuming the confounders are fully observed and the multiple imputation
	point estimators $\hat{\tau}_{\mi}$. Table \ref{t:check_cong} presents
	the simulations results of the variances of the full sample point
	estimators and the MI point estimators and the covariance between
	$\hat{\tau}_{\mi}-\hat{\tau}_{n}$ and $\hat{\tau}_{n}$. The covariance
	is significantly negative for the IPW and the matching estimators.
	Rubin's variance estimator overestimates the variances of the IPW
	estimator and matching estimator. As a consequence, MI is not congenial
	for the IPW and matching estimators. Thus, the congeniality condition
	required for MI can be quite restrictive for general ACE estimation.
	
	\begin{table}
		\centering \caption{Simulation results of the full sample point estimators and MI point
			estimators based on $5,000$ simulated data sets}
		\label{t:check_cong} %
		\begin{tabular}{l|c|c|c|c}
			\hline 
			Method $\hat{\tau}_{n}$  & $\var(\hat{\tau}_{n})$  & $\var(\hat{\tau}_{\mi})$  & $\var(\hat{\tau}_{\mi}-\hat{\tau}_{n})$  & $\cov(\hat{\tau}_{\mi}-\hat{\tau}_{n},\hat{\tau}_{n})$\tabularnewline
			& $(\times10^{4})$  & $(\times10^{4})$  & $(\times10^{4})$  & $(\times10^{4})$\tabularnewline
			\hline 
			Regression  & 24  & 35  & 11  & 0 \tabularnewline
			IPW  & 62  & 66  & 22  & -9 \tabularnewline
			AIPW  & 25  & 36  & 12  & 0\tabularnewline
			matching  & 30  & 38  & 15  & -4 \tabularnewline
			\hline 
		\end{tabular}
	\end{table}

	\section{A Martingale Representation of the MI Estimators of Causal Effects\label{sec:Martingale}}
	
	\subsection{A novel martingale representation}
	
\textcolor{black}{Based on the unified linear form of the full
	sample estimator as in (\ref{eq:linear form}) or (\ref{eq:non-ral}),
	we will express the MI estimator in a general form as 
	\begin{eqnarray}
	\hat{\tau}_{\mi}-\tau & = & \frac{1}{m}\sum_{j=1}^{m}(\hat{\tau}^{(j)}-\tau)=\frac{1}{nm}\sum_{i=1}^{n}\sum_{j=1}^{m}\psi(L_{i}^{*(j)})+o_{\pr}(n^{-1/2}),\label{eq:key formula}
	\end{eqnarray}
	where $L_{i}^{*(j)}=(L_{R_{i},i},L_{\overline{R}_{i},i}^{*(j)})$,
	and $o_{\pr}(n^{-1/2})$ is due to (\ref{eq:linear form}) or 
	\begin{eqnarray}
	\hat{\tau}_{\mi}-\tau & = & \frac{1}{m}\sum_{j=1}^{m}(\hat{\tau}^{(j)}-\tau)=\frac{1}{nm}\sum_{i=1}^{n}\sum_{j=1}^{m}\psi_{i}(\mathbf{L}^{*(j)})+o_{\pr}(n^{-1/2}),\label{eq:key formula-1}
	\end{eqnarray}
	where }\textcolor{black}{$\mathbf{L}^{*(j)}=(L_{1}^{*(j)},\ldots,L_{n}^{*(j)})$}\textcolor{black}{,
	and $o_{\pr}(n^{-1/2})$ is due to (\ref{eq:non-ral}). In the following,
	we will elucidate our framework with (\ref{eq:key formula}), and the same exposition applies to (\ref{eq:key formula-1})
	by replacing}\textcolor{black}{{} $\psi(L_{i})$ by $\psi_{i}(\mathbf{L})$
	and $L_{i}^{*(j)}$ by $\mathbf{L}^{*(j)}$. }
	
	To express (\ref{eq:key formula}) further, it is important to understand
	the properties of the posterior distribution and the imputed values
	$L_{i}^{*(j)}$. Using the Bernstein-von Mises theorem (\citealp{van1998asymptotic};
	Chapter 10), under{{} the regularity conditions described
		in Assumption \ref{asump:consistency}}, conditioned on the observed
	data, the posterior distribution $p(\theta\mid\mathbf{Z}_{\obs})$
	converges to a normal distribution with mean $\hat{\theta}$ and variance
	$n^{-1}\mathcal{I}_{\mathrm{obs}}^{-1}$ almost surely, where $\hat{\theta}$
	is the maximum likelihood estimator (MLE) of $\theta_{0}$ and $\mathcal{I}_{\mathrm{obs}}^{-1}$
	is the inverse of the Fisher information matrix. Let $S(\theta;L,R)$
	be the score function of $\theta$. In the presence of missing data,
	define the mean score function $\bar{S}(\theta_{0};Z_{\obs,i})=\E\{S(\theta_{0};L_{i},R_{i})\mid Z_{\obs,i},\theta_{0}\}$.
	
	The MLE $\hat{\theta}$ can be viewed as the solution to the mean
	score equation $\sum_{i=1}^{n}\bar{S}(\theta;Z_{\obs,i})=0$. Under{{}
		the regularity conditions described in Assumption \ref{asump:consistency}},
	we can then express $\hat{\theta}-\theta_{0}=n^{-1}\mathcal{I}_{\mathrm{obs}}^{-1}\sum_{i=1}^{n}\bar{S}(\theta_{0};Z_{\obs,i})+o_{\pr}(n^{-1/2})$.
	It is insightful to write (\ref{eq:key formula}) as 
	\begin{eqnarray}
	\hat{\tau}_{\mi}-\tau & = & \frac{1}{nm}\sum_{i=1}^{n}\sum_{j=1}^{m}\left[\psi(L_{i}^{*(j)})-\E\{\psi(L_{i})\mid \mathbf{Z}_{\obs},\hat{\theta}\}\right]\nonumber \\
	&  & +\frac{1}{nm}\sum_{i=1}^{n}\sum_{j=1}^{m}\E\{\psi(L_{i})\mid \mathbf{Z}_{\obs},\hat{\theta}\}+o_{\pr}(n^{-1/2}),\label{eq:decomp1}
	\end{eqnarray}
where we recall $\mathbf{Z}_{\obs}=(Z_{\obs,1},\ldots,Z_{\obs,n})$.
	Now, by a Taylor expansion of $\E\{\psi(L_{i})\mid \mathbf{Z}_{\obs},\hat{\theta}\}$
	around the true value $\theta_{0}$, 
	\begin{align}
	\hat{\tau}_{\mi}-\tau & =\frac{1}{nm}\sum_{i=1}^{n}\sum_{j=1}^{m}\left[\psi(L_{i}^{*(j)})-\E\{\psi(L_{i})\mid \mathbf{Z}_{\obs},\hat{\theta}\}\right]\nonumber \\
	& +\frac{1}{nm}\sum_{i=1}^{n}\sum_{j=1}^{m}\left[\E\{\psi(L_{i})\mid \mathbf{Z}_{\obs},\theta_{0}\}+\Gamma\mathcal{I}_{\mathrm{obs}}^{-1}\bar{S}(\theta_{0};Z_{\obs,i})\right]+o_{\pr}(n^{-1/2}),\label{eq:decomp}
	\end{align}
	where $\Gamma=\E\big[\E\{\psi(L_{i})S(\theta_{0};L_{i},R_{i})\mid \mathbf{Z}_{\obs},\theta_{0}\}-\E\{\psi(L_{i})\mid \mathbf{Z}_{\obs},\theta_{0}\}\bar{S}(\theta_{0};Z_{\obs,i})\big]^{\T}.$
	
	Based on (\ref{eq:decomp}), we can write 
	\begin{equation}
	n^{1/2}(\hat{\tau}_{\mi}-\tau)=\sum_{k=1}^{n+nm}\xi_{n,k}+o_{\pr}(n^{-1/2}),\label{eq:martingale}
	\end{equation}
	where 
	\[
	\xi_{n,k}=\begin{cases}
	\frac{1}{n^{1/2}}\left[\E\{\psi(L_{i})\mid \mathbf{Z}_{\obs},\theta_{0}\}+\Gamma\mathcal{I}_{\mathrm{obs}}^{-1}\bar{S}(\theta_{0};Z_{\obs,i})\right], & \text{if }k=i,\\
	\frac{1}{n^{1/2}m}\left[\psi(L_{i}^{*(j)})-\E\{\psi(L_{i})\mid \mathbf{Z}_{\obs},\hat{\theta}\}\right], & \text{if }k=n+(i-1)m+j,
	\end{cases}
	\]
	where $i=1,\ldots,n$ and  $j =1,\ldots,m$. For the decomposition in (\ref{eq:martingale}), the first $n$ terms
	of $\xi_{n,k}$ contribute to the variability of $\hat{\tau}_{\mi}$
	because of the unknown parameters, and the rest $nm$ terms of $\xi_{n,k}$
	contribute to the variability of $\hat{\tau}_{\mi}$ because of the
	imputations given the parameter values, reflecting the sequential
	MI procedure.
	
	We discuss the mean properties of $\xi_{n,k}$ in order to create
	suitable $\sigma$-fields in the martingale presentation. For $k=i,$
	where $i=1,\ldots,n,$ we have 
	\begin{eqnarray}
	\E(\xi_{n,k}) & = & \frac{1}{n^{1/2}}\E\left[\E\{\psi(L_{i})\mid \mathbf{Z}_{\obs},\theta_{0}\}+\Gamma\mathcal{I}_{\mathrm{obs}}^{-1}\bar{S}(\theta_{0};Z_{\obs,i})\right]\nonumber \\
	& = & \frac{1}{n^{1/2}}\E\{\psi(L_{i})\}+\frac{1}{n^{1/2}}\Gamma\mathcal{I}_{\mathrm{obs}}^{-1}\E\{\bar{S}(\theta_{0};Z_{\obs,i})\}=0,\label{eq:mtg1}
	\end{eqnarray}
	where $\E\{\psi(L_{i})\}=0$ and $\E\{\bar{S}(\theta_{0};Z_{\obs,i})\}=0$
	are due to the mean zero property of the influence function and the
	mean score function. For $k=n+(i-1)m+j,$ where $i=1,\ldots,n$ and  $j =1,\ldots,m$, we have
	\begin{multline}
	\E(\xi_{n,k}\mid \mathbf{Z}_{\obs})=\frac{1}{n^{1/2}m}\E\left[\psi(L_{i}^{*(j)})-\E\{\psi(L_{i})\mid \mathbf{Z}_{\obs},\hat{\theta}\}\mid \mathbf{Z}_{\obs}\right]\\
	=\frac{1}{n^{1/2}m}\left[\E\{\psi(L_{i}^{*(j)})\mid \mathbf{Z}_{\obs}\}-\E\{\psi(L_{i})\mid \mathbf{Z}_{\obs},\hat{\theta}\}\right]=0,\label{eq:mtg2}
	\end{multline}
	where the last equality follows because given $\mathbf{Z}_{\obs}$,
	the posterior predictive distribution of $L_{i}^{*(j)}$ follows the
	distribution $f(L_{i}\mid \mathbf{Z}_{\obs};\hat{\theta})$ by the Bernstein-von
	Mises theorem (\citealp{van1998asymptotic}; Chapter 10). Consider
	the $\sigma$-fields  	$\F_{n,k}=	\sigma\{\mathbb{N}\},  \text{if }k=i$ with $\mathbb{N}$ being the null set and $\F_{n,k}=	\sigma\{\mathbf{Z}_{\obs}\},  \text{if }k=n+(i-1)m+j,$
	where $i=1,\ldots,n$ and  $j =1,\ldots,m$.
	Therefore, by (\ref{eq:mtg1}) and (\ref{eq:mtg2}), 
	\[
	\left\{ \sum_{i=1}^{k}\xi_{n,i},\F_{n,k},1\leq k\leq n(1+m)\right\} \ \text{is a martingale for each }n\geq1.
	\]
	Equation (\ref{eq:decomp}) is a martingale representation of the MI estimator by expressing the MI estimator in terms of a series of random variables that have mean zero conditional on the
	sigma algebra generated from the preceding variables. This martingale representation is used to construct the bootstap replicate for variance estimation.

	\subsection{Wild bootstrap for the MI estimator}
	
	Invoked by the martingale representation, we propose the wild bootstrap procedure (\citealp{wu1986jackknife}; \citealp{liu1988bootstrap}), which provides valid variance estimation and inference of the linear statistic for martingale
	difference arrays based on the martingale central limit theory, to estimate the variance of $\hat{\tau}_{\mi}$. 
	\begin{description}
		\item [{Step$\ $1.}] Sample $u_{k}$, for $k=1,\ldots,n+nm$, to satisfy
		that $\E(u_{k}\mid\mathbf{Z}_{\obs})=0$, $\E(u_{k}^{2}\mid\mathbf{Z}_{\obs})=1$
		and $\E(u_{k}^{4}\mid\mathbf{Z}_{\obs})<\infty$. 
		\item [{Step$\ $2.}] Compute the bootstrap replicate as $T^{*}=n^{-1/2}\sum_{k=1}^{n+nm}\hat{\xi}_{n,k}u_{k}$,
		where 
		\[
		\hat{\xi}_{n,k}=\begin{cases}
		\frac{1}{n^{1/2}}\left[\E\{\psi(L_{i})\mid \mathbf{Z}_{\obs},\hat{\theta}\}+\hat{\Gamma}\hat{\mathcal{I}}_{\mathrm{obs}}^{-1}\bar{S}(\hat{\theta};Z_{\obs,i})\right], & \text{if }k=i,\\
		\frac{1}{n^{1/2}m}\left[\psi(L_{i}^{*(j)})-\E\{\psi(L_{i})\mid \mathbf{Z}_{\obs},\hat{\theta}\}\right], & \text{if }k=n+(i-1)m+j,
		\end{cases}
		\]where $i=1,\ldots,n$ and  $j =1,\ldots,m$.
		\item [{Step$\ $3.}] Repeat Step $1$--Step 2 $B$ times, and estimate
		the variance of $\hat{\tau}_{\mi}$ by the sample variance of the
		$B$ copies of $T^{*}$. 
	\end{description}
	\begin{remark}\label{rem:weight}
		
		There are many choices for generating $u_{k}$, such as the standard
		normal distribution, Mammen's two point distribution \citep{mammen1993bootstrap}
		\[
		u_{k}=\begin{cases}
		\frac{1-5^{1/2}}{2}, & \text{with probability }\frac{1+5^{-1/2}}{2},\\
		\frac{5^{1/2}+1}{2}, & \text{with probability }\frac{1-5^{-1/2}}{2},
		\end{cases}
		\]
		a simpler distribution with probability $0.5$ of being $1$ and probability
		$0.5$ of being $-1$, or {the Poisson distribution
			with parameter one re-centered at zero \citep{beyersmann2013weak}}.
		Our simulation study shows that the wild bootstrap procedure is not
		sensitive to the choice of the sampling distribution of $u_{k}$.
		In particular, one can also use the nonparametric bootstrap weights;
		that is, let $u_{k}=(nm+n)^{-1/2}(W_{k}-\overline{W})$, where $\{W_{k}:k=1,\ldots,n(m+1)\}$
		follows a multinomial distribution with $n(m+1)$ draws on $n(m+1)$
		cells with equal probability, and $\overline{W}=(nm+n)^{-1}\sum_{k=1}^{n(m+1)}W_{k}$.
		
		Several authors have used the nonparametric bootstrap to estimate
		the variance of the MI estimators. \citet{schomaker2018bootstrap}
		combined MI with bootstrap to do inference for the quantity of interest.
		However, their discussions restrict to the maximum likelihood estimators
		of model parameters and require bootstrap on top of MI, which is computationally
		intensive. Moreover, in the causal inference literature in the absence
		of missing data, \citet{abadie2008failure} has demonstrated that
		nonparametric bootstrap can not provide consistent variance estimation
		for the matching estimators of the ACE due to the non-smooth nature
		of the matching procedure. It is important to note that the proposed
		wild bootstrap procedure with the nonparametric bootstrap weights
		is different from the naive bootstrap. The martingale representation
		and the wild bootstrap procedure work for the asymptotically linear
		ACE estimators including the matching estimator.
		
	\end{remark}

\begin{remark}\label{rem:estimate}
 In Step 2, we require approximating $\xi_{n,k}$, which involves the MLE $\hat{\theta}$, the estimated observed Fisher information, and the conditional expectations taken with respect to the distribution of the missing values given the observed values. These estimators are readily available from the posterior draws or approximated by Monte Carlo integration based on the imputed values. For example, we approximate $\E\{\psi(L_{i})\mid \mathbf{Z}_{\obs},\hat{\theta}\}$ by $M^{-1}\sum_{j=1}^{M}\psi(L_{i}^{*(j)})$. Thus, the computation is not as intimidating as it appears, although it is heavier than Rubin's combining rule. However, as shown in Theorem 1, the proposed inference procedure is valid, while Rubin's method may not.
	\end{remark}
	
	We show the asymptotic validity of the above bootstrap inference method
	by the following theorem with regularity assumptions.
	
\begin{assumption}\label{asump:consistency} Suppose the standard
	conditions hold for the maximum likelihood estimator (MLE) $\hat{\theta}$
	to be $n^{1/2}$-consistent for $\theta_{0}$: 
	\begin{enumerate}
		\item $Z_{\obs,1},\dots,Z_{\obs,n}$ are independently and identically distributed
		and follow $f(z\mid\theta)$; 
		\item $\theta$ is identifiable; i.e., if $\theta\neq\theta'$, then $f(z\mid\theta)\neq f(z\mid\theta')$; 
		\item The density $f(z\mid\theta)$ have a common support (not depend on
		$\theta$); 
		\item The parameter space contains an open set of which the true parameter
		$\theta_{0}$ is an interior point.; 
		\item For every $z$ in the support, $f(z\mid\theta)$ is three times differentiable
		with respect to $\theta$, the third derivative is continuous in $\theta$,
		and $\int\partial^{3}\log f(z\mid\theta)/\partial\theta^{3}\de z<\infty$; 
		\item For any $\theta_{0}$ in the parameter space, there exists a positive
		number $c$ and a function $M(z)$ such that $\left|\partial^{3}\log f(z\mid\theta)/\partial\theta^{3}\right|\leq M(z)$
		for all $z$ in the support, $\theta_{0}-c<\theta<\theta_{0}+c$,
		with $\E_{\theta_{0}}\{M(Z)\}<\infty$. 
	\end{enumerate}
\end{assumption}

Define $\bar{\psi}(\theta;Z_{\obs,i})=\E\{\psi(L_{i})\mid Z_{\obs,i},\theta\}$.

\begin{assumption}{\label{asump:continuity}} $\bar{\psi}(\theta;\mathbf{Z}_{\obs})$,
	$\var\{\psi(L_{i})\mid \mathbf{Z}_{\obs},\theta\}$, $\bar{S}(\theta;Z_{\obs,i})$
	and $\var\{S(\theta;L_{i},R_{i})\mid Z_{\obs,i},\theta\}$ are continuous
	functions of $\theta$. \end{assumption}

\begin{assumption}{\label{asump:bound}} $\E[\{\bar{\psi}(\theta;\mathbf{Z}_{\obs})\}^{4}]<\infty$
	and $\E[\{\bar{S}(\theta;Z_{\obs,i})\}^{4}]<\infty$ for $\theta$
	in a neighborhood of $\theta_{0}$. \end{assumption}

\begin{assumption}{\label{asump:donsker}} $\{\bar{\psi}(\theta;\mathbf{Z}_{\obs}-\bar{\psi}(\theta_{0};\mathbf{Z}_{\obs})\}^{2}$
	and $\{\bar{S}(\theta;Z_{\obs,i})-\bar{S}(\theta_{0};Z_{\obs,i})\}^{2}$
	belong to a Donsker class. \end{assumption}

	Assumption \ref{asump:consistency} is the standard
	assumption in the literature to guarantee the consistency of the MLE
	\citep{van1998asymptotic}. Assumption \ref{asump:continuity} is
	imposed to guarantee sufficient smoothness on the conditional mean
	and variance functions for the influence function and the score function.
	It holds for the general estimands such as mean-type estimands and
	the commonly-used class of parametric models such as the exponential
	family. For Assumption \ref{asump:bound}, the moment conditions are
	used to invoke the central limit theory and typically hold for the
	general estimands and parametric models coupled with the bounded moment
	conditions for $L$. In practice, $L$ often has a bounded support
	and thus the bounded moment conditions are reasonable. Assumption
	\ref{asump:donsker} ensures the convergence of the empirical process
	to its limiting version \citep{kennedy2016semiparametric}. The interested
	readers can consult \citet{kennedy2016semiparametric} for details
	and examples of the Donsker class.

	\begin{theorem}\label{th:consistency} Suppose that Assumptions \ref{asump-ignorable},
		\ref{asump-overlap} and \ref{asump:consistency}-\ref{asump:donsker} hold. Suppose that
			$f(L_{\overline{R}_{i},i}\mid Z_{\obs,i};\theta)$ is correctly specified.
			Then, \textcolor{black}{for MI adopts the full sample estimator that satisfies (\ref{eq:linear form}) or (\ref{eq:non-ral})},
			we have  
		\[
		\underset{r}{\rm{sup}}\left\vert \pr(n^{1/2}T^{*}\leq r\mid\mathbf{Z}_{\obs})-\pr\{n^{1/2}(\hat{\tau}_{\mi}-\tau)\leq r\}\right\vert \overset{\pr}{\to}0,
		\]
		as $n\to\infty$. \end{theorem}
	
	We provide the proof of Theorem \ref{th:consistency} in the supplementary
	material, which draws on the martingale central limit theory \citep{hall1980martingale}
	and the asymptotic property of weighted sampling of martingale difference
	arrays \citep{pauly2011weighted}. Theorem \ref{th:consistency} indicates
	that the distribution of the wild bootstrap statistic consistently
	estimates the distribution of the MI estimator.
	
	Theorem \ref{th:consistency} requires {the imputation
		model $f(L_{\overline{R}_{i},i}\mid Z_{\obs,i};\theta)$ to be correctly
		specified (the congeniality condition of \citealp{meng1994multiple}).
		This requirement is needed not only for the consistency of the MI
		variance estimator but also for the consistency of the MI point estimator.
		Corollaries hereafter clarify the required correct imputation models
		in different scenarios. }
	
	\begin{corollary}\label{cr:mar}{For the scenario
			with confounders missing at random, the assumption that the imputation
			model $f(L_{\overline{R}_{i},i}\mid Z_{\obs,i};\theta)$ is correctly
			specified in Theorem \ref{th:consistency} implies that the outcome
			distribution $f(Y_{i}\mid X_{i},A_{i};\theta)$, the propensity score
			model $f(A_{i}\mid X_{i};\theta)$ and the confounder distribution
			$f(X_{\overline{R}_{Xi},i}\mid X_{R_{Xi},i};\theta)$ should be correctly
			specified.}
		
	\end{corollary}
	
	\section{Extensions\label{sec:Extension}}
	
	\subsection{Different causal estimands\label{subsec:Other-esimands}}
	
	Our inference framework extends to a wide class of causal estimands,
	as long as the estimand admits an asymptotically linear full sample
	estimator as in (\ref{eq:linear form}). For example, we can consider
	the average causal effects over a subset of the population \citep{crump2006moving,li2018balancing},
	including the average causal effect on the treated. We can also consider
	nonlinear causal estimands. For example, for a binary outcome, the
	log of the causal risk ratio is 
	\[
	\log\text{CRR}=\log\frac{\pr\{Y(1)=1\}}{\pr\{Y(0)=1\}}=\log\frac{\E\{Y(1)\}}{\E\{Y(0)\}},
	\]
	and the log of the causal odds ratio is 
	\[
	\log\text{COR}=\log\frac{\pr\{Y(1)=1\}/\pr\{Y(1)=0\}}{\pr\{Y(0)=1\}/\pr\{Y(0)=0\}}=\log\frac{\E\{Y(1)\}/[1-\E\{Y(1)\}]}{\E\{Y(0)\}/[1-\E\{Y(0)\}]}.
	\]
	The key insight is that under Assumptions \ref{asump-ignorable} and
	\ref{asump-overlap}, we can estimate $\E\{Y(a)\}$ with commonly-used
	estimators, denoted by $\hat{\E}\{Y(a)\}$, for $a=0,1$. We can then
	obtain and estimator for the $\log\text{CRR}$ as $\log[\hat{\E}\{Y(1)\}/\hat{\E}\{Y(0)\}]$.
	By the Taylor expansion, we can linearize these estimators and establish
	a similar linear form as (\ref{eq:linear form}), which serves as
	the basis to construct the weighted bootstrap inference.
	
	\subsection{Missingness not at random\label{sec:mnar}}
	
	If Assumption \ref{assump:MAR} fails, the missing pattern also depends
	on the missing values themselves even after controlling for the observed
	data, a scenario known as missing not at random (MNAR). In our motivating
	example discussed in Section \ref{sec:realdata}, the family poverty
	ratio is likely to be missing not at random because subjects with
	higher income may be less likely to disclose their income information
	\citep{davern2005effect}. In general, MNAR occurs frequently for
	sensitive questions regarding e.g. alcohol consumption, income, etc.
	
	Causal inference with data missing not at random is more challenging
	because the full data distribution and therefore the ACE are not identifiable
	in general. To utilize MI in causal inference with confounders MNAR,
	we require identification conditions that ensure the full data distribution
	is identifiable. For example, \citet{wang2014instrumental} introduced
	a nonresponse instrument as a sufficient condition for the identifiability
	of the observed likelihood. \citet{miao2016identifiability} investigated
	the identifiability of normal and normal mixture models with nonignorable
	missing data. \citet{yang2019causal} proposed an outcome-independence
	missingness mechanism under which the missing data mechanism is independent
	of the outcome given the treatment and confounders and established
	general identification conditions.
	
	Our proposed method can easily extend to the scenario where the confounders
	are MNAR when additional assumptions are made for identifiability
	of the full data distribution. After the identification check, we
	only need to modify the posterior predictive distribution of $X_{\overline{R}_{i},i}^{(j)}$.
	For example, following \citet{yang2019causal}, we assume that the
	missingness pattern $R$ is independent of the outcome given the treatment
	and confounders.
	
	\begin{assumption}[Outcome-independent missingness]\label{assump:MAR-1}We
		have $Y\perp\!\!\!\perp R_{X}\mid(A,X_{R_{X}},X_{\overline{R}_{X}}).$
		
	\end{assumption}
	
	{Under the regularity conditions in \citet{yang2019causal}},
	$f(A,X,Y,R_{X})$ is identifiable \citep{yang2019causal}. Then in
	Step MI-1, the posterior distribution of $X_{\overline{R}_{Xi},i}^{(j)}$
	can be decomposed to 
	\begin{multline*}
	f(X_{\overline{R}_{Xi},i}\mid A_{i},X_{R_{Xi},i},Y_{i},R_{Xi};\theta^{*(j)})\propto f(Y_{i}\mid X_{R_{Xi},i},X_{\overline{R}_{Xi},i},A_{i};\theta^{*(j)})\\
	\times f(R_{Xi}\mid X_{R_{Xi},i},X_{\overline{R}_{Xi},i},A_{i};\theta^{*(j)})f(A_{i}\mid X_{R_{Xi},i},X_{\overline{R}_{Xi},i};\theta^{*(j)})f(X_{\overline{R}_{Xi},i}\mid X_{R_{Xi},i};\theta^{*(j)}).
	\end{multline*}
	After imputation, the wild bootstrap steps remain exactly the same.
	
	\begin{corollary}\label{cr:mnar}{For the scenario
			with confounders missing not at random, the assumption that the imputation
			model $f(L_{\overline{R}_{i},i}\mid Z_{\obs,i};\theta)$ is correctly
			specified in Theorem \ref{th:consistency} implies that the outcome
			distribution $f(Y_{i}\mid X_{i},A_{i};\theta)$, the propensity score
			model $f(A_{i}\mid X_{i};\theta)$, the confounder distribution $f(X_{\overline{R}_{Xi},i}\mid X_{R_{Xi},i};\theta)$,
			and the missingness model $f(R_{Xi}\mid X_{i},A_{i};\theta)$ should
			be correctly specified.}
		
	\end{corollary}
	
	\subsection{Partially observed outcome and confounders\label{subsec:Partially-observed-outcome}}
	
	In some cases, both the outcome and the confounders are subject to
	missingness. Our framework can easily accommodate this scenario by
	adding an outcome imputation step in the MI procedure.
	
	We now introduce another missingness indicator $R_{Y}$ for $Y$;
	i.e., $R_{Y}=1$ if $Y$ is observed and $R_{Y}=0$ otherwise. In
	Step MI-1, we first generate $\theta^{*(j)}$ from the posterior distribution
	$p(\theta\mid\mathbf{Z}_{\obs})$. Then for unit $i$ with $R_{Y}=1$,
	generate $X_{\overline{R}_{Xi},i}^{*(j)}$ from $f(X_{\overline{R}_{Xi},i},\mid A_{i},X_{R_{Xi},i},Y_{i},R_{i},R_{Yi}=1;\theta^{*(j)})$;
	for unit $i$ with $R_{Y}=0$, generate $X_{\overline{R}_{Xi},i}^{*(j)}$
	and $Y_{i}^{*(j)}$ from $f(X_{\overline{R}_{Xi},i},Y_{i}\mid A_{i},X_{R_{Xi},i},R_{Xi},R_{Yi}=0;\theta^{*(j)})$
	to create the $j$the imputed data set. Then the MI estimator can
	be written in a general form with both imputed outcome and confounders
	as 
	\begin{eqnarray*}
		\hat{\tau}_{\mi}-\tau & = & \frac{1}{nm}\sum_{i=1}^{n}\sum_{j=1}^{m}\psi(A_{i},X_{i}^{*(j)},Y_{i}^{*(j)})+o_{\pr}(1).
	\end{eqnarray*}
	Accordingly, the martingale difference arrays in the wild bootstrap
	procedure can be written as 
	\[
	\hat{\xi}_{n,k}=\begin{cases}
	\frac{1}{n^{1/2}}\left[\E\{\psi(A_{i},X_{i},Y_{i})\mid \mathbf{Z}_{\obs},\hat{\theta}\}+\hat{\Gamma}\hat{\mathcal{I}}_{\mathrm{obs}}^{-1}\bar{S}(\hat{\theta};Z_{\obs,i})\right], & \text{if }k=i,\\
	\frac{1}{n^{1/2}m}\left[\psi(A_{i},X_{i}^{*(j)},Y_{i}^{*(j)})-\E\{\psi(A_{i},X_{i},Y_{i})\mid \mathbf{Z}_{\obs},\hat{\theta}\}\right], & \text{if }k=n+(i-1)m+j,
	\end{cases}
	\]where $i=1,\ldots,n$ and  $j =1,\ldots,m$.
	Other steps in the MI and wild bootstrap procedures remain the same
	as described for the scenario when only confounders have missing values.
	
	\begin{corollary}{For the scenario where both the
			outcome and the confounders are subject to missingness, the assumption
			that the imputation model $f(L_{\overline{R}_{i},i}\mid Z_{\obs,i};\theta)$
			is correctly specified in Theorem \ref{th:consistency} implies Corollary
			\ref{cr:mar} under MAR and Corollary \ref{cr:mnar} under MNAR. }
		
	\end{corollary}
	
	\section{Simulation study\label{sec:sim}}
	
	We conduct simulation studies to evaluate the finite sample performance
	of the proposed inference when MI adopts different full sample estimators
	including the outcome regression, IPW, AIPW and matching estimators.
	
	For each sample, the confounder $X=(X_{[1]},X_{[2]})$ are sampled
	from a multivariate normal distribution with mean $(0,0)$, variance
	$(1,1)$ and a correlation coefficient $0.2$. The potential outcomes
	follow $Y(0)=2+3X_{[1]}+2X_{[2]}+\epsilon(0)$ and $Y(1)=1+2X_{[1]}+X_{[2]}+\epsilon(1)$,
	where $\epsilon(0)\sim\N(0,\sigma_{0}^{2})$, $\epsilon(1)\sim\N(0,\sigma_{1}^{2})$
	with $\sigma_{0}=\sigma_{1}=1$, and $\epsilon(0)$ and $\epsilon(1)$
	are independent. So the true value of ACE is $\tau=-1$. We generate
	the treatment indicator $A$ from Bernoulli$\{\pi_{A}(X)\}$ and $\pi_{A}(X)=P(A=1\mid X)=\Phi(-0.2+0.3X_{[1]}+0.4X_{[2]})$,
	where $\Phi(\cdot)$ is the cumulative density function for the standard
	normal distribution. In the sample, we assume $A$ and $X_{[1]}$
	are fully observed, but $X_{[2]}$ and $Y$ can be partially observed
	with the missing indicators $R_{[2]}$ and $R_{Y}$, respectively.
	We consider four scenarios: 
	\begin{description}
		\item [{(a)}] $X_{[2]}$ is missing at random; i.e., its missingness depends
		only on the observed data. Let $R_{[2]}\sim\text{Bernoulli}\{\pi_{R1}(A,X_{[1]},Y)\}$,
		where $\pi_{R1}(A,X_{[1]},Y)=\Phi(-0.1+0.1A+0.5X_{[1]}+0.2Y)$ with
		the missingness rate being about $45\%$. Moreover, the inference
		procedure assumes the correct missingness mechanism; 
		\item [{(b)}] $X_{[2]}$ is missing not at random; i.e., its missingness
		depends on unobserved data. Let $R_{[2]}\sim\text{Bernoulli}\{\pi_{R2}(A,X_{[1]},X_{[2]})\}$,
		where $\pi_{R2}(A,X_{[1]},X_{[2]})=\Phi(0.2+1X_{[2]})$ with the missingness
		rate being about $45\%$. Moreover, the inference procedure assumes
		the correct missingness mechanism; 
		\item [{(c)}] $X_{[2]}$ is missing not at random as in scenario (b); but
		the inference procedure assumes an incorrect missingness at random
		mechanism; 
		\item [{(d)}] both $X_{[2]}$ and $Y$ are missing not at random, with
		the missingness indicators $R_{[2]}$ and $R_{Y},$ respectively.
		Let $R_{[2]}\sim\text{Bernoulli}\{\pi_{R}(X_{[2]})\}$, where $\pi_{R}(X_{[2]})=\Phi(0.8+1X_{[2]})$
		with the missingness rate being about $30\%$. Let $R_{Y}\sim\text{Bernoulli}\{\pi_{Y}(A,X)\}$,
		where $\pi_{Y}(A,X)=\Phi(1+0.2A+0.5X_{[1]}+0.5X_{[2]})$ with the
		missingness rate being about $20\%$. 
	\end{description}
	We generate $5,000$ Monte Carlo samples with size $n=3000$ for each
	scenario. In MI, the missing data mechanism is specified according
	to the above scenarios and other components of the distribution are
	correctly specified. We use non-informative priors for parameters.
	Suppose that the prior distribution for each coefficient in the outcome
	model, the propensity score model and the missing indicator model
	is $\N(0,100)$; the prior distribution for the variance parameters
	$\sigma_{0}$ and $\sigma_{1}$ in the outcome regression model is
	Gamma$(0.01,0.01)$; the prior distribution for the mean of $X$ is
	$(0,0)$; the prior distribution for the variance covariance matrix
	of X is $I_{2}$, where $I_{2}$ is the 2-dimensional identity matrix. \textcolor{black}{More details about priors and posterior sampling are provided in the supplementary material.}
	We consider three sizes of multiple imputation with $m=5,10\text{ or }100$.
	To generate the posterior samples of the missing values $X_{\overline{R}}^{*(j)}$,
	we use Gibbs sampling with $5,000$ iterations, discard first $2,000$
	burn-in samples, and randomly choose $m$ posterior samples from the
	remaining $3,000$ draws. For each imputed data set, we calculate
	the full sample point estimators and variance estimators of the ACE
	using outcome regression, IPW, AIPW and matching, and then use Rubin's
	method to get the corresponding MI estimators $\hat{\tau}_{\mi}$
	and Rubin's variance estimators $\hat{V}_{\mi}$. For the matching
	estimator, we set the number of matches as $M=1$.
	
	We compare the standard MI inference and the proposed bootstrap inference.
	For the standard MI inference, the $100(1-\alpha)\%$ confidence intervals
	are calculated as $(\hat{\tau}_{\mi}-t_{\nu,1-\alpha/2}\hat{V}_{\mi}^{1/2},\hat{\tau}_{\mi}+t_{\nu,1-\alpha/2}\hat{V}_{\mi}^{1/2})$,
	where $t_{\nu,1-\alpha/2}$ is the $100(1-\alpha/2)\%$ quantile of
	the $t$ distribution with degree of freedom $\nu=(m-1)\lambda^{-2}$
	with $\lambda=(1+m^{-1})B_{m}/\{W_{m}+(1+m^{-1}B_{m})\}$. For the
	proposed bootstrap procedure, we use $B=1,000$, generate the weights
	$\mu_{k}$ from the Mammen's two point distribution as suggested in
	Remark \ref{rem:weight}, and calculate the variance estimate $\hat{V}_{\text{BS}}$.
	The corresponding $100(1-\alpha)\%$ confidence interval are estimated
	using two different methods: (i) quantile-based confidence interval
	$(\hat{\tau}_{\mi}-q_{1-\alpha/2}^{*},\hat{\tau}_{\mi}-q_{\alpha/2}^{*})$,
	where $q_{1-\alpha/2}^{*}$ and $q_{\alpha/2}^{*}$ are the $(1-\alpha/2)$th
	and $(\alpha/2)$th quantiles of $T^{*}$; (ii) the Wald-type confidence
	interval $(\hat{\tau}_{\mi}-z_{1-\alpha/2}\hat{V}_{\text{BS}}^{1/2},\hat{\tau}_{\mi}+z_{1-\alpha/2}\hat{V}_{\text{BS}}^{1/2})$,
	where $z_{1-\alpha/2}$ is the $(1-\alpha/2)$th quantile of the standard
	normal distribution.
	
	We assess the performance in terms of the relative bias of the variance
	estimator and the coverage rate of confidence intervals. The relative
	bias of the variance estimators are calculated as $\{\E(\hat{V}_{\mi})-\var(\hat{\tau}_{\mi})\}/\var(\hat{\tau}_{\mi})\times100\%$
	and $\{\E(\hat{V}_{\text{BS}})-\var(\hat{\tau}_{\mi})\}/\var(\hat{\tau}_{\mi})\times100\%$
	correspondingly. The coverage rate of the $100(1-\alpha)\%$ confidence
	intervals is estimated by the percentage of the Monte Carlo samples
	for which the confidence intervals contain the true value.
	
	Tables \ref{t:MAR}--\ref{t:y_mis} present the simulation results
	for the four scenarios. When the imputation model is correctly specified
	as in scenarios (a), (b) and (d), the MI point estimator has small
	biases for all full sample estimators. Also, as $m$ increases, the
	variance of the MI point estimator becomes smaller, suggesting that
	using more imputations can help improving the efficiency of the MI
	estimator. Across difference choices of m, the relative bias of the proposed variance estimator stays small. The accuracy of the proposed variance estimator is less sensitive to the choice of m. Rubin's variance estimator is unbiased for the outcome
	regression estimator and the AIPW estimator; however, it overestimates
	the variances of the IPW estimator and the matching estimator e.g.
	by as high as $29.7\%$ and $20.1\%$ in scenario (a). Due to variance
	overestimation, the coverage rate of Rubin's method exceeds
	the nominal level for the IPW and Matching estimators, all exceeding $96\%$ and some reaching $97.3\%$. In contrast, our proposed wild
	bootstrap procedure for variance estimation is unbiased for all four
	ACE estimators, and therefore the coverage rate of the confidence
	intervals based on our proposed wild bootstrap method is close to
	the nominal level.  Moreover, the proposed method is not sensitive
	to the number of imputations $m$ and the choice of quantile-based
	or Wald-type confidence interval. However, in scenario (c) when the
	true missing data mechanism is missingness not at random while the
	inference procedure assumes missingness at random, the MI point estimator
	has large biases and all the confidence intervals have poor coverage
	rates; see Table \ref{t:trtMNAR}.
	
	There are other methods developed for multiple imputation inference. For example, \citet{xie2017dissecting} proposed a doubling variance approach for more conservative variance estimation when Rubin's method underestimates the variance. However, it will further overestimates the variance of MI estimators in our simulation settings so that the performance is even worse than Rubin's method. \citet{meng1992performing} and \citet{chan2022multiple} proposed likelihood ratio based procedure for multiply-imputed data inference. However, this procedure is not easily implemented for the variance and confidence interval construction for the treatment effect estimation.
	
	\begin{table}
		\centering \caption{Simulation results: point estimate (Monte Carlo mean of point estimates),
			true variance (Monte Carlo variance of point estimates), relative
			bias of the variance estimator, coverage and mean width of interval
			estimate using Rubin's method and the proposed wild bootstrap method
			under scenario (a) with missingness at random}
		\label{t:MAR} \resizebox{\textwidth}{!}{ %
			\begin{tabular}{lccccccccccc}
				\hline 
				Method $\hat{\tau}_{n}$  & $m$  & Point est  & True var  & \multicolumn{2}{c}{Relative Bias} & \multicolumn{3}{c}{Coverage (\%)} & \multicolumn{3}{c}{Mean width ($\times10^{2}$)}\tabularnewline
				&  & $(\times10)$  & $(\times10^{4})$  & \multicolumn{2}{c}{(\%)} & \multicolumn{3}{c}{for 95\% CI} & \multicolumn{3}{c}{for 95\% CI}\tabularnewline
				&  &  &  & Rubin  & BS  & Rubin  & \multicolumn{2}{c}{BS} & Rubin  & \multicolumn{2}{c}{BS}\tabularnewline
				&  &  &  &  &  &  & Quantile  & Wald  &  & Quantile  & Wald\tabularnewline
				\hline 
				& 5  & -10.0  & 35.8  & -2.1  & 1.9  & 94.3  & 94.9  & 95.4  & 23.9  & 23.6  & 24.1 \tabularnewline
				Regression  & 10  & -10.0  & 34.9  & -1.9  & 3.7  & 94.6  & 95.3  & 95.8  & 23.1  & 23.6  & 24.0\tabularnewline
				& 100  & -10.0  & 33.8  & -1.4  & 5.6  & 94.8  & 95.6  & 95.9  & 22.6  & 23.4  & 23.9\tabularnewline
				\hline 
				& 5  & -10.0  & 68.0  & \textbf{25.8}  & \textbf{-0.3}  & 96.0  & 93.9  & 94.7  & 35.6  & 31.1  & 31.9\tabularnewline
				IPW  & 10  & -10.0  & 66.3  & \textbf{27.4}  & \textbf{0.3}  & 96.3  & 94.2  & 94.6  & 34.9  & 30.8  & 31.6\tabularnewline
				& 100  & -10.0  & 64.4  & \textbf{29.7}  & \textbf{1.2}  & 96.3  & 94.2  & 94.7  & 34.4  & 30.4  & 31.3\tabularnewline
				\hline 
				& 5  & -10.0  & 36.6  & 3.0  & -3.9  & 95.2  & 94.4  & 94.9  & 24.8  & 23.2  & 23.7\tabularnewline
				AIPW  & 10  & -10.0  & 35.7  & 3.0  & -2.7  & 94.9  & 94.5  & 95.0  & 24.0  & 23.1  & 23.5\tabularnewline
				& 100  & -10.0  & 34.6  & 3.7  & -1.1  & 95.3  & 94.7  & 95.3  & 23.5  & 22.9  & 23.4\tabularnewline
				\hline 
				& 5  & -10.0  & 39.1  & \textbf{18.2}  & \textbf{-4.5}  & 96.5  & 94.4  & 95.0  & 27.5  & 23.9  & 24.4\tabularnewline
				Matching  & 10  & -10.0  & 37.8  & \textbf{18.7}  & \textbf{-3.5}  & 96.5  & 94.5  & 95.1  & 26.6  & 23.7  & 24.2\tabularnewline
				& 100  & -10.0  & 36.4  & \textbf{20.1}  & \textbf{-2.1}  & 96.9  & 94.4  & 95.0  & 26.0  & 23.4  & 23.9\tabularnewline
				\hline 
		\end{tabular}} 
	\end{table}
	
	\begin{table}
		\centering \caption{Simulation results under scenario (b) with missingness not at random}
		\label{t:MNAR} \resizebox{\textwidth}{!}{ %
			\begin{tabular}{lccccccccccc}
				\hline 
				Method $\hat{\tau}_{n}$  & $m$  & Point est  & True var  & \multicolumn{2}{c}{Relative Bias} & \multicolumn{3}{c}{Coverage (\%)} & \multicolumn{3}{c}{Mean width ($\times10^{2}$)}\tabularnewline
				&  & $(\times10)$  & $(\times10^{4})$  & \multicolumn{2}{c}{(\%)} & \multicolumn{3}{c}{for 95\% CI} & \multicolumn{3}{c}{for 95\% CI}\tabularnewline
				&  &  &  & Rubin  & BS  & Rubin  & \multicolumn{2}{c}{BS} & Rubin  & \multicolumn{2}{c}{BS}\tabularnewline
				&  &  &  &  &  &  & Quantile  & Wald  &  & Quantile  & Wald\tabularnewline
				\hline 
				& 5  & -10.0  & 34.5  & -0.5  & 2.8  & 94.6  & 95.2  & 95.7  & 23.6  & 23.3  & 23.8\tabularnewline
				Regression  & 10  & -10.0  & 33.6  & 0.9  & 4.4  & 94.8  & 95.4  & 95.7  & 22.9  & 23.2  & 23.7\tabularnewline
				& 100  & -10.0  & 32.9  & -0.1  & 5.6  & 94.8  & 95.5  & 96.0  & 22.5  & 23.1  & 23.6\tabularnewline
				\hline 
				& 5  & -10.0  & 67.5  & \textbf{28.0}  & \textbf{0.3}  & 96.4  & 94.5  & 94.8  & 35.7  & 30.9  & 31.7\tabularnewline
				IPW  & 10  & -10.0  & 65.6  & \textbf{30.6}  & \textbf{1.3}  & 96.7  & 94.6  & 95.0  & 35.0  & 30.6  & 31.4\tabularnewline
				& 100  & -10.0  & 64.2  & \textbf{29.8}  & \textbf{1.4}  & 96.7  & 94.7  & 95.0  & 34.5  & 30.4  & 31.2\tabularnewline
				\hline 
				& 5  & -10.0  & 35.5  & 5.0  & -2.3  & 95.2  & 94.8  & 95.2  & 24.6  & 23.1  & 23.5\tabularnewline
				AIPW  & 10  & -10.0  & 34.5  & 5.6  & -0.7  & 95.5  & 94.9  & 95.5  & 23.9  & 22.9  & 23.4\tabularnewline
				& 100  & -10.0  & 33.6  & 5.7  & -0.5  & 95.5  & 95.1  & 95.4  & 23.4  & 22.8  & 23.2\tabularnewline
				\hline 
				& 5  & -10.0  & 38.0  & \textbf{21.0}  & \textbf{-3.5}  & 96.9  & 94.8  & 95.4  & 27.5  & 23.7  & 24.2\tabularnewline
				Matching  & 10  & -10.0  & 36.7  & \textbf{21.8}  & \textbf{-2.1}  & 96.9  & 95.0  & 95.5  & 26.5  & 23.5  & 24.0\tabularnewline
				& 100  & -10.0  & 35.6  & \textbf{22.4}  & \textbf{-1.1}  & 97.0  & 94.9  & 95.3  & 25.9  & 23.2  & 23.7\tabularnewline
				\hline 
		\end{tabular}} 
	\end{table}
	
	\begin{table}
		\centering \caption{Simulation results under scenario (c) when the true missing mechanism
			is missing not at random but missingness at random is assumed}
		\label{t:trtMNAR} \resizebox{\textwidth}{!}{ %
			\begin{tabular}{llcccccccccc}
				\hline 
				Method $\hat{\tau}_{n}$  & $m$  & Point est  & True var  & \multicolumn{2}{c}{Relative Bias} & \multicolumn{3}{c}{Coverage (\%)} & \multicolumn{3}{c}{Mean width ($\times10^{2}$)}\tabularnewline
				&  & $(\times10)$  & $(\times10^{4})$  & \multicolumn{2}{c}{(\%)} & \multicolumn{3}{c}{for 95\% CI} & \multicolumn{3}{c}{for 95\% CI}\tabularnewline
				&  &  &  & Rubin  & BS  & Rubin  & \multicolumn{2}{c}{BS} & Rubin  & \multicolumn{2}{c}{BS}\tabularnewline
				&  &  &  &  &  &  & Quantile  & Wald  &  & Quantile  & Wald\tabularnewline
				\hline 
				& 5  & -11.5  & 34.6  & 1.7  & 10.9  & 27.2  & 29.1  & 30.2  & 23.7  & 24.3  & 24.8\tabularnewline
				Regression  & 10  & -11.5  & 33.8  & 1.8  & 12.3  & 25.6  & 24.1  & 24.6  & 23.2  & 24.1  & 24.6\tabularnewline
				& 100  & -11.5  & 33.2  & 1.4  & 13.0  & 23.9  & 27.9  & 28.9  & 22.8  & 24.0  & 24.5\tabularnewline
				\hline 
				& 5  & -12.0  & 130.1  & \textbf{31.5 }  & \textbf{1.1 }  & 66.1  & 54.5  & 53.7  & 46.3  & 39.1  & 40.5\tabularnewline
				IPW  & 10  & -12.0  & 127.8  & \textbf{31.3 }  & \textbf{-1.4 }  & 64.9  & 53.1  & 51.9  & 45.6  & 38.6  & 40.0\tabularnewline
				& 100  & -12.0  & 126.4  & \textbf{33.3 }  & \textbf{-1.8 }  & 64.7  & 52.1  & 50.9  & 45.4  & 38.2  & 39.6\tabularnewline
				\hline 
				& 5  & -11.5  & 36.3  & 6.0  & -0.7  & 31.0  & 27.5  & 28.6  & 24.7  & 23.5  & 24.0\tabularnewline
				AIPW  & 10  & -11.5  & 35.5  & 5.8  & 0.2  & 29.0  & 26.5  & 27.8  & 24.1  & 23.3  & 23.8\tabularnewline
				& 100  & -11.5  & 34.9  & 5.5  & 0.5  & 27.6  & 26.3  & 27.4  & 23.8  & 23.2  & 23.7\tabularnewline
				\hline 
				& 5  & -11.6  & 38.7  & \textbf{26.2 }  & \textbf{-1.3 }  & 40.9  & 29.4  & 30.8  & 28.1  & 24.2  & 24.7\tabularnewline
				Matching  & 10  & -11.6  & 37.5  & \textbf{26.6 }  & \textbf{-0.5 }  & 38.4  & 27.8  & 29.1  & 27.3  & 23.9  & 24.4\tabularnewline
				& 100  & -11.6  & 36.6  & \textbf{26.7 }  & \textbf{-0.2 }  & 36.5  & 27.2  & 28.6  & 26.7  & 23.6  & 24.1\tabularnewline
				\hline 
		\end{tabular}} 
	\end{table}
	
	\begin{table}
		\centering \caption{Simulation results under scenario (d) where both the outcome and confounders
			are missing and missing not at random is assumed}
		\label{t:y_mis} \resizebox{\textwidth}{!}{ %
			\begin{tabular}{lccccccccccc}
				\hline 
				Method $\hat{\tau}_{n}$  & $m$  & Point est  & True var  & \multicolumn{2}{c}{Relative Bias} & \multicolumn{3}{c}{Coverage (\%)} & \multicolumn{3}{c}{Mean width ($\times10^{2}$)}\tabularnewline
				&  & $(\times10)$  & $(\times10^{4})$  & \multicolumn{2}{c}{(\%)} & \multicolumn{3}{c}{for 95\% CI} & \multicolumn{3}{c}{for 95\% CI}\tabularnewline
				&  &  &  & Rubin  & BS  & Rubin  & \multicolumn{2}{c}{BS} & Rubin  & \multicolumn{2}{c}{BS}\tabularnewline
				&  &  &  &  &  &  & Quantile  & Wald  &  & Quantile  & Wald\tabularnewline
				\hline 
				& 5  & -10.0  & 35.6  & -2.4  & -1.5  & 94.6  & 94.7  & 95.2  & 23.7  & 23.2  & 23.7 \tabularnewline
				Regression  & 10  & -10.0  & 34.3  & -0.9  & 0.7  & 94.9  & 95.0  & 95.7  & 23.1  & 23.0  & 23.5 \tabularnewline
				& 100  & -10.0  & 33.4  & -0.5  & 2.2  & 95.0  & 95.3  & 95.7  & 22.6  & 22.9  & 23.4 \tabularnewline
				\hline 
				& 5  & -10.0  & 68.5  & \textbf{28.6}  & \textbf{-2.7}  & 96.3  & 94.2  & 94.7  & 36.6  & 30.8  & 31.6\tabularnewline
				IPW  & 10  & -10.0  & 65.9  & \textbf{32.7}  & \textbf{-0.8}  & 96.7  & 94.5  & 94.9  & 35.6  & 30.4  & 31.3\tabularnewline
				& 100  & -10.0  & 64.0  & \textbf{34.3}  & \textbf{-0.2}  & 97.3  & 94.5  & 95.1  & 35.2  & 30.1  & 30.9\tabularnewline
				\hline 
				& 5  & -10.0  & 36.5  & 7.3  & -3.9  & 95.5  & 94.4  & 94.9  & 25.4  & 23.2  & 23.7\tabularnewline
				AIPW  & 10  & -10.0  & 34.9  & 9.7  & -1.3  & 96.1  & 94.6  & 95.4  & 24.5  & 23.0  & 23.5\tabularnewline
				& 100  & -10.0  & 33.8  & 10.2  & 0.1  & 96.1  & 94.9  & 95.3  & 23.9  & 22.8  & 23.3\tabularnewline
				\hline 
				& 5  & -10.0  & 39.5  & \textbf{18.5}  & \textbf{-4.7}  & 96.6  & 94.1  & 94.6  & 27.8  & 24.0  & 24.5\tabularnewline
				Matching  & 10  & -10.0  & 37.7  & \textbf{21.4}  & \textbf{-2.6}  & 97.1  & 94.5  & 95.0  & 26.8  & 23.7  & 24.2\tabularnewline
				& 100  & -10.0  & 36.5  & \textbf{22.1}  & \textbf{-1.5}  & 97.2  & 94.8  & 95.6  & 26.2  & 23.5  & 24.0\tabularnewline
				\hline 
		\end{tabular}} 
	\end{table}

	\section{An application\label{sec:realdata}}
	
	We apply our method to a dataset from the 2015-2016 U.S. National
	Health and Nutrition Examination Survey to estimate the ACE of education
	on general health satisfaction. The general health satisfaction outcome
	($Y$) is fully observed with a lower value indicating better satisfaction.
	A sample of $4,845$ individuals is divided into two groups: one ($76\%$)
	with at least high school education, denoted as $A=1$, and the other
	one ($24\%$) with education level lower than high school, denoted
	as $A=0$. The covariates $X$ consist of four categorical variables
	including age, race, gender, marital status, and one continuous variable
	family poverty ratio which is truncated at $0$ and $5$. The family
	poverty ratio has about $10\%$ missing values. The other four covariates
	are fully observed.
	
	{The general health satisfaction outcome ($Y$) is
		an ordinal variable with distinct values $1,2,3,4,5$. We introduce
		a latent continuous variable $Y^{*}$ to link the ordinal outcome
		to the continuous space with support $(-\infty,+\infty)$: 
		\[
		Y=\begin{cases}
		1 & \text{if }Y^{*}<1,\\{}
		[Y^{*}] & \text{if }1\leq Y^{*}\leq5,\\
		5 & \text{if }Y^{*}>5.
		\end{cases}
		\]
		where $[\cdot]$ represents rounding to the nearest integer. Since
		the family poverty ratio $X_{[1]}$ is a continuous variable truncated
		at $0$ and $5$, we introduce another latent variable $X_{[1]}^{*}$
		to link the recorded truncated family poverty ratio values to the
		full continuous space $(-\infty,+\infty)$: 
		\[
		X_{[1]}=\begin{cases}
		0 & \text{if }X_{[1]}^{*}<0,\\
		X_{[1]}^{*} & \text{if }0\leq X_{[1]}^{*}\leq5,\\
		5 & \text{if }X_{[1]}^{*}>5.
		\end{cases}
		\]
		Accordingly, let $X^{*}$ include the latent family poverty ratio
		variable $X_{[1]}^{*}$ and the other four variables. }To facilitate
	imputation and estimation, we assume the latent outcome $Y^{*}$ follows
	a linear regression model, i.e., $Y^{*}(a)=X^{*\T}\beta_{a}+\epsilon(a)$,
	where $\epsilon(a)\sim\N(0,\sigma_{a}^{2})$ for $a=0,1$. The treatment
	indicator follows Bernoulli$\{\pi_{A}(X^{*})\}$ with $\pi_{A}(X^{*})=\Phi(X^{*\T}\alpha)$.
	The missing indicator follows Bernoulli$\{\pi_{R}(X^{*},A)\}$ with
	$\pi_{R}(X^{*},A)=\Phi\{(X^{*},A)^{\T}\gamma\}$, under which the
	missingness of the family poverty ratio probably depend on the missing
	values themselves but not the outcome variable (i.e., Assumption \ref{assump:MAR-1}).
	{Also, we assume the latent family poverty ratio
		follows a linear regression model with the other covariates, i.e.,
		$X_{\overline{R}_{X}}^{*}=X_{R_{X}}\eta+\epsilon_{X}$, where
		$X_{\overline{R}_{X}}^{*}=X_{[1]}^{*}$ represents the latent family
		poverty ratio and $X_{R}$ represents the other four covariates, $\epsilon_{X}\sim\N(0,\sigma_{X}^{2})$.
		We have conducted model diagnoses in the supplementary material and
		the diagnosis plots show that the proposed model fits the data well.
		Given the outcome model and the covariate model, the missing values
		of the family poverty ratio can be imputed by $f(X_{\overline{R}_{X}}^{*}\mid A,X_{R_{X}},Y,R_{X};\theta^{*(j)})\propto f(Y^{*}\mid X^{*},A;\theta^{*(j)})f(R_{X}\mid X^{*},A;\theta^{*(j)})f(A\mid X^{*};\theta^{*(j)})f(X_{\overline{R}_{X}}^{*}\mid X_{R_{X}};\theta^{*(j)})$
		given each posterior sample of the parameters $\theta^{*(j)}$.} \textcolor{black}{More details about priors and posterior sampling are provided in the supplementary material.}
	
	For each imputed dataset, we consider the full sample point estimators
	of the ACE using outcome regression, IPW, AIPW, and matching based
	on propensity score to reduce the dimensionality of the matching variable
	\citep{abadie2016matching}. We compare Rubin's variance estimator
	and the proposed wild bootstrap variance estimator. Table \ref{t:realdata}
	shows that education has a significantly positive effect on the general
	health satisfaction. The variances for the IPW estimator and the matching
	estimator estimated by Rubin's method are larger than the variances
	estimated by the wild bootstrap method, while the two methods give
	similar results for the regression estimator and the AIPW estimator.
	This suggests Rubin's method works well for the regression estimator
	and the AIPW estimator but might overestimate the variances of the
	IPW and matching estimators, which is consistent with our observations
	in the simulation studies.
	
	\begin{table}
		\centering \caption{Result for the ACE of education on general health satisfaction: point
			estimates, the variance of point estimators, and 95\% confidence interval
			estimated using Rubin's method and proposed wild bootstrap method.}
		\label{t:realdata}
		
		\begin{tabular}{lccccc}
			\hline 
			&  & \multicolumn{2}{c}{Rubin} & \multicolumn{2}{c}{BS}\tabularnewline
			Method  & Point est  & Var est  & 95\% CI  & Var est  & 95\% CI\tabularnewline
			&  & $(\times10^{4})$  &  & $(\times10^{4})$  & Wald\tabularnewline
			\hline 
			Regression  & -0.36  & 19  & (-0.45,-0.27)  & 19  & (-0.45,-0.27)\tabularnewline
			IPW  & -0.25  & 65  & (-0.41,-0.10)  & 54  & (-0.40,-0.11)\tabularnewline
			AIPW  & -0.27  & 32  & (-0.38,-0.16)  & 31  & (-0.38,-0.16)\tabularnewline
			Matching  & -0.25  & 40  & (-0.37,-0.12)  & 28  & (-0.35,-0.14)\tabularnewline
			\hline 
		\end{tabular}
	\end{table}

	\section{Conclusion\label{sec:conclude}}
	
	This paper establishes a unified inference framework for multiple
	imputation using martingale which invokes the wild bootstrap inference
	for consistent variance estimation. Our framework allows a wide class
	of asymptotically linear full sample estimators. We demonstrate its
	utility in estimating the ACE with missing values. The simulation
	results indicate the good finite sample performance of the proposed
	method when MI adopts different full sample estimators including the
	outcome regression, IPW, AIPW, and matching estimators. Our framework
	works well when the missing mechanism is either MAR or MNAR. 
	
	Our framework can also be extended in the following directions. First,
	multiple imputation was originated for survey data, which often contain
	design weights (or sample weights) to account for sample selection.
	If sampling weights are non-informative, the sample data follow the
	population model, and therefore the imputation can be done by ignoring
	sampling weights; whereas, if sampling weights are informative, the
	sample data distribution is different from the population model, and
	therefore imputation must take into account sampling weights. The
	full Bayesian imputation is difficult (if not impossible) to implement
	in this case. To mitigate this problem, \citet{kim2017note} and \citet{wang2018approximate}
	proposed an approximate Bayesian computation technique, which can
	be used for multiple imputation in complex sampling. It would be interesting
	to extend the martingale representation to this setting in our future
	work. Second, in the current work, we assume that the imputer's model
	and the analyst's model are the same and are correctly specified.
	\citet{xie2017dissecting} argued that the uncongeniality of the imputer's
	model and the analyst's model is the rule but not an exception. Their
	findings suggest that even both models are correctly specified, if
	the imputation model is more saturate than the analysis model, the
	standard MI inference may be invalid. In future work, we will extend
	our framework to this setting for consistent inference allowing uncongeniality.
	
	\section*{Acknowledgment}
	
	Yang is partially supported by the NSF DMS 1811245, NIH 1R01AG066883,
	and 1R01ES031651.
	
	\section*{Supplementary materials}
	
	The online supplementary material contains common estimators of the ACE and their influence functions, proofs, \textcolor{black}{the priors and MCMC details for the simulation study and application}, model diagnosis in the application, and the R code that
	implements the proposed method is available at \url{https://github.com/qianguan/miATE}.

\par

\bibhang=1.7pc
\bibsep=2pt
\fontsize{9}{14pt plus.8pt minus .6pt}\selectfont
\renewcommand\bibname{\large \bf References}
\expandafter\ifx\csname
natexlab\endcsname\relax\def\natexlab#1{#1}\fi
\expandafter\ifx\csname url\endcsname\relax
  \def\url#1{\texttt{#1}}\fi
\expandafter\ifx\csname urlprefix\endcsname\relax\def\urlprefix{URL}\fi

\bibliographystyle{chicago}      
\bibliography{ci}               

\vskip .65cm

\noindent
Department of Statistics, North Carolina State University
\vskip 2pt
\noindent
E-mail: qguan2@ncsu.edu, syang24@ncsu.edu

\end{document}